\documentclass[a4paper]{article}

\usepackage[english]{babel}
\usepackage[utf8x]{inputenc}

\usepackage{booktabs}
\usepackage{tabu}
\usepackage[T1]{fontenc}

\usepackage[a4paper,top=3cm,bottom=2cm,left=3cm,right=3cm,marginparwidth=1.75cm]{geometry}

\usepackage{amsmath}
\usepackage{graphicx}
\usepackage[colorinlistoftodos]{todonotes}
\usepackage[colorlinks=true, allcolors=blue]{hyperref}

\title{Analysis and Evaluation of the Inequality of the Spatial Distribution of Medical Resources in Jinan}
\author{Research Proposal \\ ZHU Shengkun}
\date{}

\begin{document}
\maketitle

\section*{Summary of the Proposal}
This article will analyze the inequality and evaluation of the spatial distribution of medical resources in Jinan. The research will be carried out from the following four aspects: analysis of existing medical resource allocation and distribution characteristics, medical resource accessibility analysis, inequality evaluation and optimization layout analysis. The article will use G2SFCA/M2SFCA Model, Spatial Clustering Analysis and HRAD.

\section*{Background}

With economic development and improvement of residents’ living quality, their demand for urban public service facilities is continually increasing. As an important component of urban public service facilities, medical facilities provide residents with necessary medical services. But there is an increasing inconsistency between the multi-level medical and health demand and the apparent of medical facilities. Therefore, the spatial distribution, allocation and equalization of medical facilities have practical significance.

\section*{Goal and Objectives}

This article starts with the four aspects of the spatial distribution characteristics, spatial accessibility, inequality evaluation, and layout optimization of the existing medical resource allocation in Jinan. It also provides a scientific basis for the optimization of the layout of medical facilities and the rational allocation of resources in Jinan. Based on the analysis of the article, targeted countermeasures and suggestions are put forward for layout optimization.

\section*{Technical Route}
\clearpage

\begin{figure}[h]
	
	\centering
	
	\includegraphics[scale=0.2315]{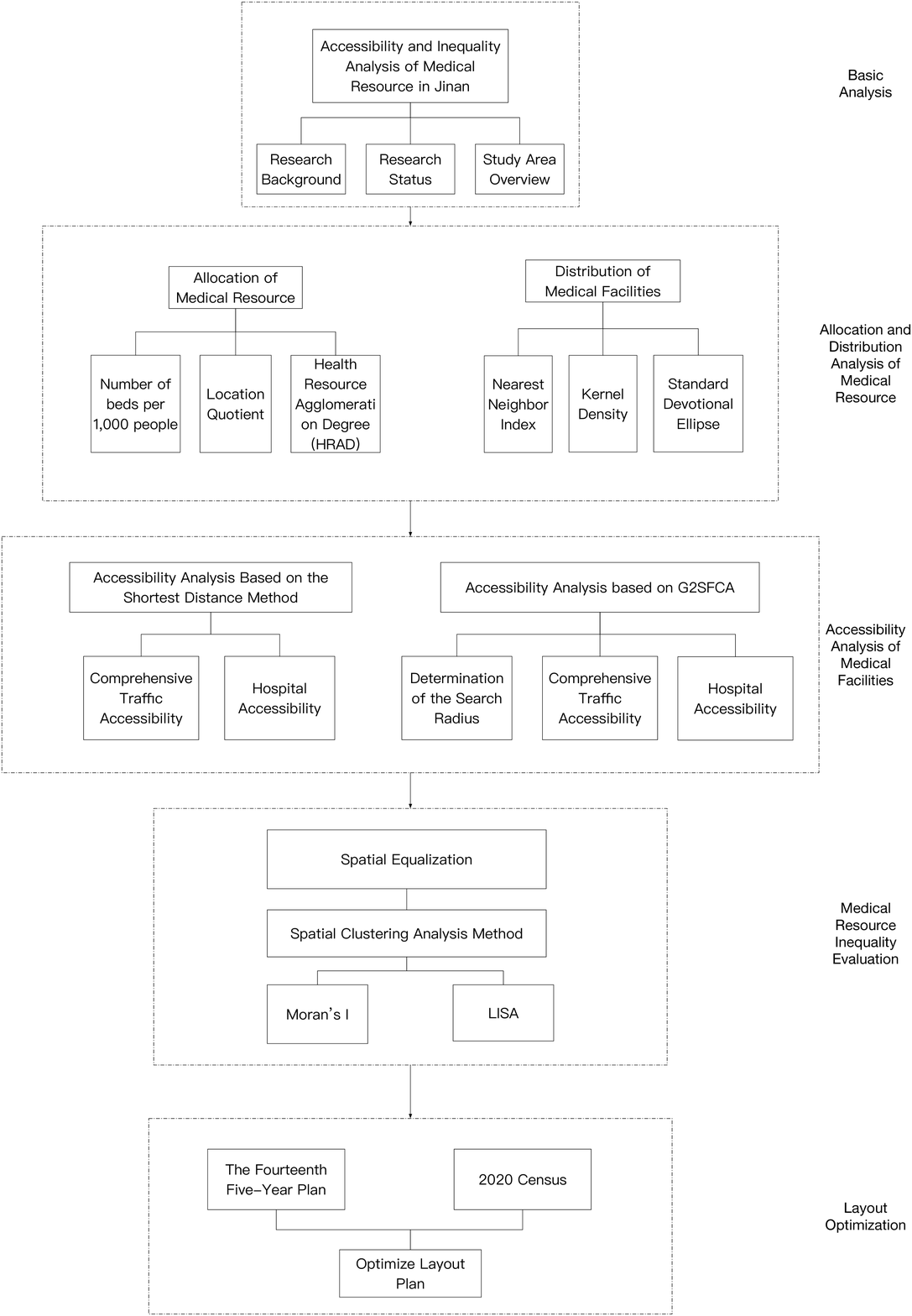}
	
	\caption{Technical Route}
	
	\label{fig:label}
	
\end{figure}
\clearpage

\section*{Methods}
\subsection*{2SFCA}
Aiming at the deficiencies of the gravity model, Luo and Wang\cite{2003Measures} proposed the 2-Step Floating Catchment Area Method (2SFCA method), and then Luo and Qi \cite{2009An} applied the distance attenuation function in different traffic time zones, and proposed Enhanced 2-Step Floating Catchment Area method, (E2SFCA method). Subsequently, scholars successively proposed Modified 2-Step Floating Catchment Area method (M2SFCA method), Variable Catchment sies for the 2-Step Floating Catchment Area method (V2SFCA method) , 3-Step Floating Catchment Area method (3SFCA method)\cite{2012Variable}. Because now researchers more often use the E2SFCA method and the M2SFCA method.

\begin{itemize}
\item E2SFCA
\end{itemize}
The E2SFCA method consists of two steps. In the calculation of each search area, regardless of the first step or the second step, it is divided into multiple sub-regions, and each sub-region will be assigned a different weight according to the Gaussian function.\cite{2007Immigration}

The E2SFCA method is similar to the continuous gravity model. This method subdivides each search area, solves the problem of regional differences and makes the results of the E2SFCA method intuitive and easy to use.

In addition, when dividing sub-areas, this method uses a reasonable approximation, because in actual situations, people seldom care about the difference in a few minutes when seeking medical services.\cite{2009An} However, the E2SFCA method still has the disadvantages of the FCA method. Compared with the Gravity-based Model, the FCA method artificially divides the accessible and inaccessible medical service points in a sense.\cite{2007Immigration} In addition, $\beta$ -related issues remain unresolved.\cite{2010Measuring,2012A}

\begin{itemize}
	\item M2SFCA
\end{itemize}
What cannot be ignored is that all FCA laws implicitly assume that the geographic location of medical service providers is in an optimally configured state and can meet all the medical service needs of people in the region. However, in reality, it is impossible for the medical and health service system to achieve an optimal configuration.\cite{2013Spatial} Therefore, some scholars proposed the M2SFCA method to solve this dilemma.

If the structure between the residential area and the medical service point is well configured, the difference between the spatial accessibility results calculated by the E2SFCA method and the M2SFCA method is very small. On the contrary, the spatial accessibility results calculated by the E2SFCA method and the M2SFCA method will have a large deviation.

\begin{itemize}
	\item G2SFCA
\end{itemize}
Wang (2012) \cite{Wang2012Measurement} reviews various refinements to the original 2SFCA method and proposes the Generalized 2SFCA method as a framework to synthesize all:

$${A_i} = \sum\nolimits_{j = 1}^n {[{S_j}f({d_{ij}})/\sum\nolimits_{k = 1}^m {({D_k}f({d_{kj}}))} ]} $$

The method is convenient to implement in a GIS environment (Wang 2015 \cite{0Planning}). Its result can be intuitively interpreted as the supplydemand ratio (e.g. doctors per person; or doctors per 1,000 people if A is inflated 1,000 times), and a larger value indicates better access (Wang 2015 \cite{Tu2015Quantitative})



\subsection*{Spatial Equalization}
\begin{itemize}
	\item Moran's I
\end{itemize}
Moran's I is one of the most widely used global indexes. It usually uses a single attribute to reflect whether the neighboring areas in the study area are similar, different or independent of each other, to determine whether the attribute value has a clustering feature in space, and then to reflect Its degree of equalization.\cite{que}
\begin{itemize}
	\item LISA
\end{itemize}
It is difficult for Moran's I to detect the spatial location of clusters and regional correlation patterns, and even if the research area is not significantly correlated in the global clustering detection, there may still be local correlation phenomena. Therefore, in practical applications, Moran's I and LISA are usually combined to determine and identify spatial correlation and correlation patterns. The LISA model can not only calculate the degree of association between each spatial unit and a certain attribute of its neighboring units, but also reveal the similarity between the attribute value of the research unit and its neighboring spatial units.\cite{lijian}




\begin{itemize}
	\item HRAD
\end{itemize}
Evaluation of health resource allocation equity is one of the main contents of health system research. In the paper, the concept of agglomeration degree is introduced into health resource allocation evaluation, and the health resource agglomeration degree (HRAD) method is put forward. Through taking the demographic and geographical factor into consideration, the feasibility of HRAD in health resource allocation equity evaluation field is explored.\cite{yuan}

$$HRA{D_i} = \frac{{(H{R_i}/H{R_n}) \times 100\% }}{{({A_i}/{A_n}) \times 100\% }} = \frac{{H{R_i}/{A_i}}}{{H{R_n}/{A_n}}}$$

${HRA{D_i} = 1}$, which means absolutely equality. ${HRA{D_i} > 1}$, which means relatively fair, ${HRA{D_i} < 1}$, which means unfair.

\subsection*{Layout Optimization Model}
In this section, the article will combine the 14th Five-Year Plan of National Economic and Social Development of Jinan City and the data of the 7th Census in 2020, and through the population changes in the past ten years and the outline of the future five-year plan, the article will address the inequitable allocation of medical resources analyzed in the article to adjust and layout the resources with a view to achieving the maximum utilization and equity of resources.
\section*{Future}
In the future, we can study the spatial distribution and inequality of medical resources for a specific disease, for example, chronic diseases such as diabetes and heart disease, and acute diseases such as colds and fever. At the same time, the scope of research may not be limited to the urban area of Jinan, such as the inequality caused by factors such as urban-rural differences and differences in medical resources of different levels of hospitals.

\bibliographystyle{plain}
\bibliography{refs}

\end{document}